\documentclass[preprint,superscriptaddress,showkeys,showpacs,
tightenlines,fleqn,nofootinbib,prd,byrevtex]{revtex4} 
\usepackage{graphicx}
\begin{document}      
\preprint{INHA-NTG-16/2008}
\preprint{YITP-08-73}
\title{Pion and kaon structures from the instanton vacuum}
\author{Hyun-Chul Kim}
\email[E-mail:]{hchkim@inha.ac.kr}
\affiliation{Department of Physics, Inha University, Incheon 402-751,
Republic of Korea}
\author{Seung-il Nam}
\email[E-mail:]{sinam@yukawa.kyoto-u.ac.jp}
\affiliation{Yukawa Institute for Theoretical Physics (YITP), 
Kyoto University, Kyoto 606-8502, Japan} 
\author{Hyun-Ah Choi}
\email[E-mail:]{hachoi@pusan.ac.kr} 
\affiliation{Department of Physics, Pusan National University,\\ 
San 30 Jangjeon-dong, Geumjung-gu, Busan, 609-735, Korea}
\date{September, 2008}

\begin{abstract}
In this presentation, we briefly review recent investigations on pion 
and kaon structures from the instanton vacuum.  Starting from the
low-energy QCD partition function, we have computed the
Gasser-Leutwyler low-energy constants, electromagnetic form factors of
the pion and kaon, semileptonic decay form factors of the kaon, and
light-cone distribution amplitues of the pion and kaon.  The results
are in good agreement with the experimental and empirical data. 
\end{abstract}

\maketitle

\section{Introduction}	
It is of great importance to understand structures of the pion and
kaon, since they are regarded as the pesudo-Goldstone 
bosons arising from the spontaneous breakdown of chiral symmetry
(SB$\chi$S).  Though there are many theoretical frameworks to study
the pion and kaon such as chiral perturbation theory, QCD sum rules,
and lattice QCD, the instanton-vacuum approach~(see, for example,
recent reviews~\cite{Diakonov:2002fq,Schafer:1996wv}) has certain
virtues: First, it can be directly related to nonperturbative QCD,
realizing well the mechanism of the SB$\chi$S with the quark zero
modes.  Second, it has only two parameters: The average instanton size 
$\bar \rho\approx \frac{1}{3}\, \mathrm{fm}$ and average
inter-instanton distance $\bar R\approx 1\, \mathrm{fm}$. Thus, there
is basically no free parameter in this approach.  Third, the
model has a natural normalization point that is given by the
average size of instantons ($\rho^{-1}\approx 0.6\, \mathrm{GeV}$). It
allows us to calculate hadronic observables and compare the results
with the experimental data with the renormalization group employed. 

The staring point is the following gauge-invariant low-energy
effective chiral action:  
\begin{equation}
\mathcal{S}_{\rm eff}[\mathcal{M}^{\alpha},V,m]  = -{\rm Sp}_{c,f,\gamma} \ln
\left[i\rlap{/}{D}+im + i\sqrt{M (iD)}  U^{\gamma_5}\sqrt{M
  (iD)}\right]. 
\label{eq:eca}  
\end{equation}
On can refer to Ref.~\cite{Nam:2007gf} for the description of
Eq.~(\ref{eq:eca}) in detail.  
\section{Low-energy constants}
The SU(3) effective chiral Lagrangian to order $\mathcal{O}(p^4)$ in
the large $N_c$ and in the chiral limit is expressed as
\begin{equation}
{\cal L }^{(4)}=L_1 \left \langle \partial_{\mu} U^{\dagger}
  \partial_{\mu}U\right\rangle^2  + L_2 \left \langle \partial_{\mu}
U^{\dagger} \partial_{\nu} U \right \rangle^2  
+ L_3 \left \langle \partial_{\mu} U^{\dagger} \partial_{\mu} U
  \partial_{\nu} U^{\dagger} \partial_{\nu} U \right \rangle,
\end{equation}
where $L_i$ denote the Gasser-Leutwyler low-energy constants.
Utilizing the derivative expansion of Eq.~(\ref{eq:eca}), we can
determine the $L_i$~\cite{Choi:2003cz}.  The results are listed in
Table 1. The explanation for the abbreviation can be found in
Ref.~\cite{Choi:2003cz}.  The results are in good agreement with the
empirical values by Gasser and Leutwyler~\cite{Gasser:1983yg}.
\begin{table}[ht]
\caption{The low energy constants $L_1$, $L_2$, $L_3$.}
{\begin{tabular}{@{}cccccc@{}} \\ \toprule
& $M_0$(MeV) & $\Lambda $(MeV) &$L_1 ( \times 10^{-3})$&
$L_2 ( \times 10^{-3})$ &$L_3 ( \times 10^{-3})$ \\ \hline 
DP  & $350$ & $611.7$& $0.82$ & $1.63$ & $-3.09$ \\ 
Dipole & $350$ & $611.2$ & $0.82$ & $1.63$ & $-2.97$ \\  
Gaussian & $350$ & $627.4$ & $0.81$ & $1.62$ & $-2.88$ \\ 
GL & & & $0.9\pm 0.3$ & $1.7\pm 0.7$ & $-4.4 \pm 2.5$ \\
\botrule
\end{tabular} }
\end{table} 

The combination of the low-energy constant is related to the upper
bound of the sigma-meson mass:
\begin{equation}
M_{\sigma} < 665[1+0.44 \Delta +0.33 \Delta ^{2} 
+ {\cal O}(\Delta ^{3})] {\rm MeV},
\label{Eq:Delta}  
\end{equation}
where $\Delta=- (2L_2 +L_3)/L_2$ which can be determined by the
$\pi\pi$ scattering length. The $\Delta$ is ranged from $-1.03$ to
$-0.243$, depending on the type of the form factor.  As a result, the
upper bound of the sigma-meson mass lies in the range of $610\sim 640$
MeV, which is very different from almost all other models in which
$\Delta$ turns out to be positive.  
\section{Electromagnetic form factors of the pion and kaon}
The electromagnetic form (EM) factors of the pion and kaon are defined as
follows: 
\begin{equation}
 \label{eq:emff}
\langle \mathcal{M} (P_f)|j_{\mu}^{\rm EM}(0) |\mathcal{M} (P_i)\rangle =
(P_i+P_f)_{\mu}F_{\mathcal{M}} (q^2).  
\end{equation}
The notation of Eq.~(\ref{eq:emff}) can be found in
Ref.~\cite{Nam:2007gf}.  The results of the EM form factors of the
pion and kaon are drawn in Fig.~1 (see Ref.~\cite{Nam:2007gf} for the
explanation of the abbreviation). 
\begin{figure}[ht]
\includegraphics[width=7cm]{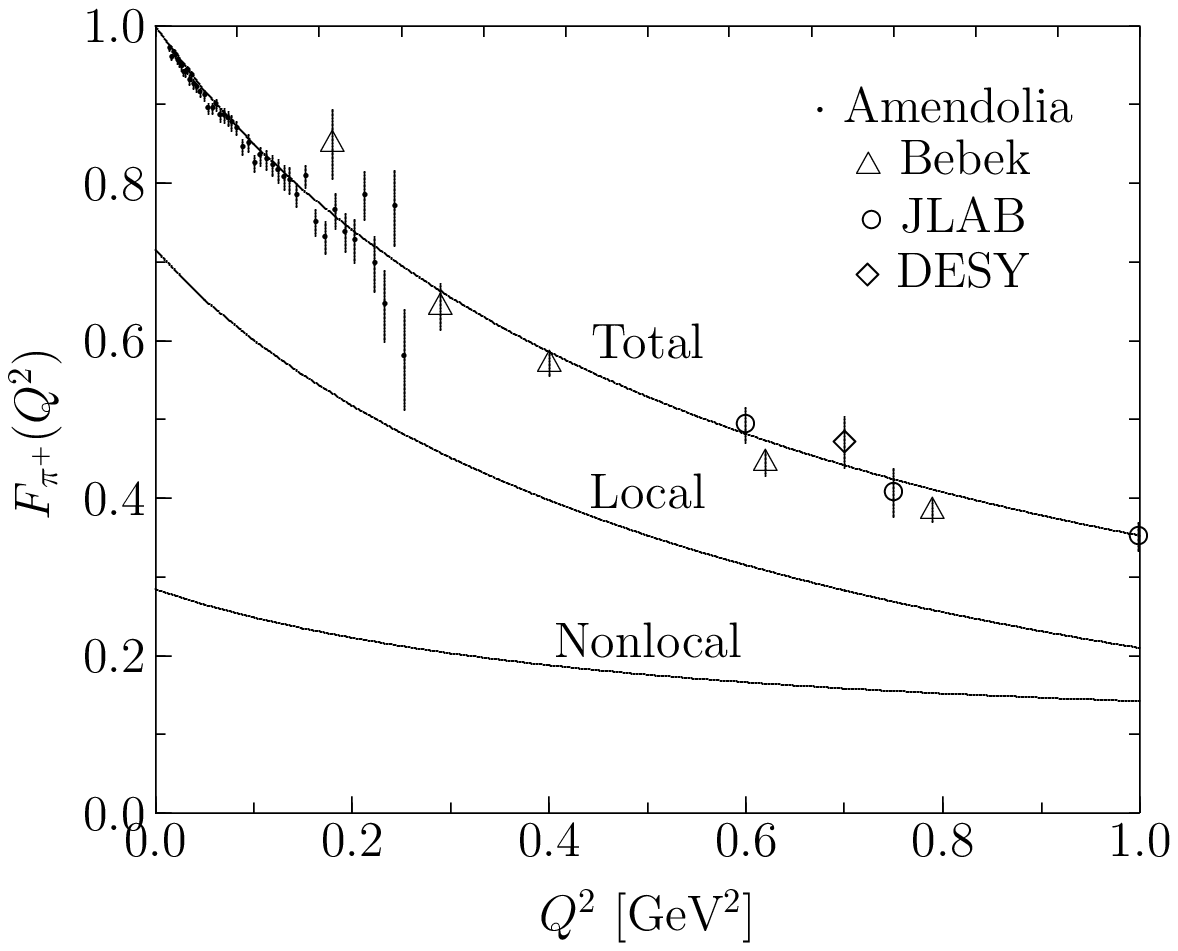}
\includegraphics[width=7.4cm]{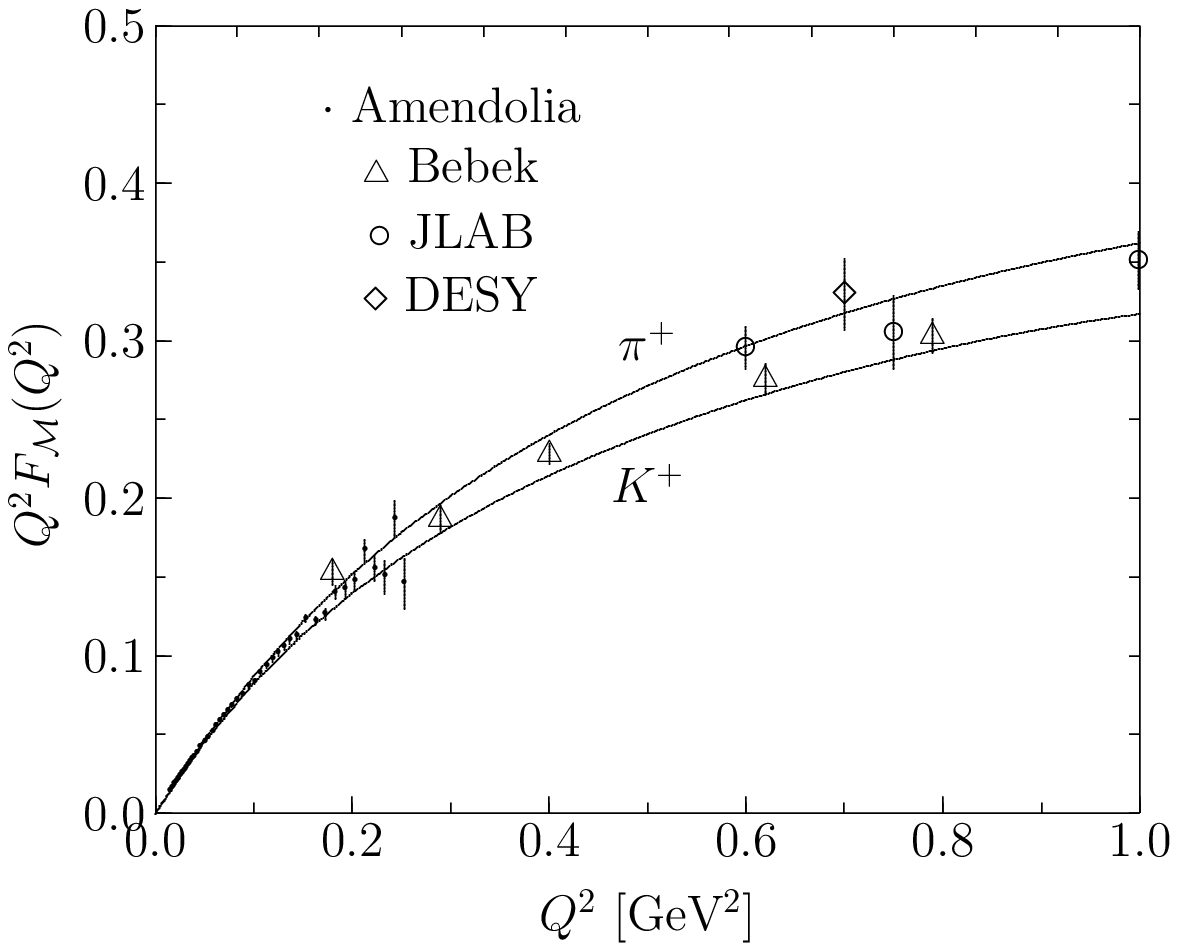}
\vspace*{8pt}
\caption{The results of the electromagnetic form factors of the pion
  and kaon~\protect\label{fig1}}
\end{figure}
The EM charge radii of the pion and kaon are: $\langle
  r^2\rangle_{\pi^+}^{1/2}=0.675$ fm  and $\langle
  r^2\rangle_{K^+}^{1/2}=0.731$ fm (experimental data: $\langle
r^2\rangle_{\pi^+}^{1/2}=0.672\pm 0.008$ fm and $\langle
  r^2\rangle_{K^+}^{1/2}=0.560\pm 0.031$ fm), respectively. The
deviation of the kaon charge radius from the experment is due to the
fact that the   $1/N_c$ corrections are absent in the present
approach. 
\section{Semileptonic decays of the kaon}
The corresponding matrix element of the kaon semileptonic decay form
factors is written as 
\begin{equation}
\langle \pi(p_\pi)|\bar{\psi}\gamma_{\mu}\lambda^{4+i5}
\psi|K(p_K)\rangle
\;=\;(p_K+p_{\pi})_{\mu}f_{l+}(t)+(p_K-p_{\pi})_{\mu}f_{l-}(t).    
\label{eq:F3}
\end{equation}
The detailed description for Eq.~(\ref{eq:F3}) can be found in
Ref.~\cite{Nam:2007fx}.  The $t$-dependence of the form factor is
shown in the left panel of Fig. 2.  The results are in qualitative
agreement with the data. The dimensionless slope parameter
$\lambda_{l+}$ is obtained as $3.03\times 10^{-2}$ (experimet:
$(2.96\pm0.05)\times10^{-2}$).  
\section{Light-cone distribution amplitudes of the pion and kaon}
The gauge invariance of Eq.~(\ref{eq:eca}) plays an important role in
deriving the light-cone distribution amplitudes (DA) of the
pion~\cite{Nam:2006sx}.  In the right panel of Fig.~2, we draw the
Gegenbauer moments relevant for the pion DA. The ellipses denote the
Schmedding-Yakovlev analysis~\cite{Schmedding:1999ap}.  The results
lie within the $2\sigma$ ellipsis.  One can consult
Ref.~\cite{Nam:2006sx} for details.
\begin{figure}[ht]
\centering
\includegraphics[width=7cm]{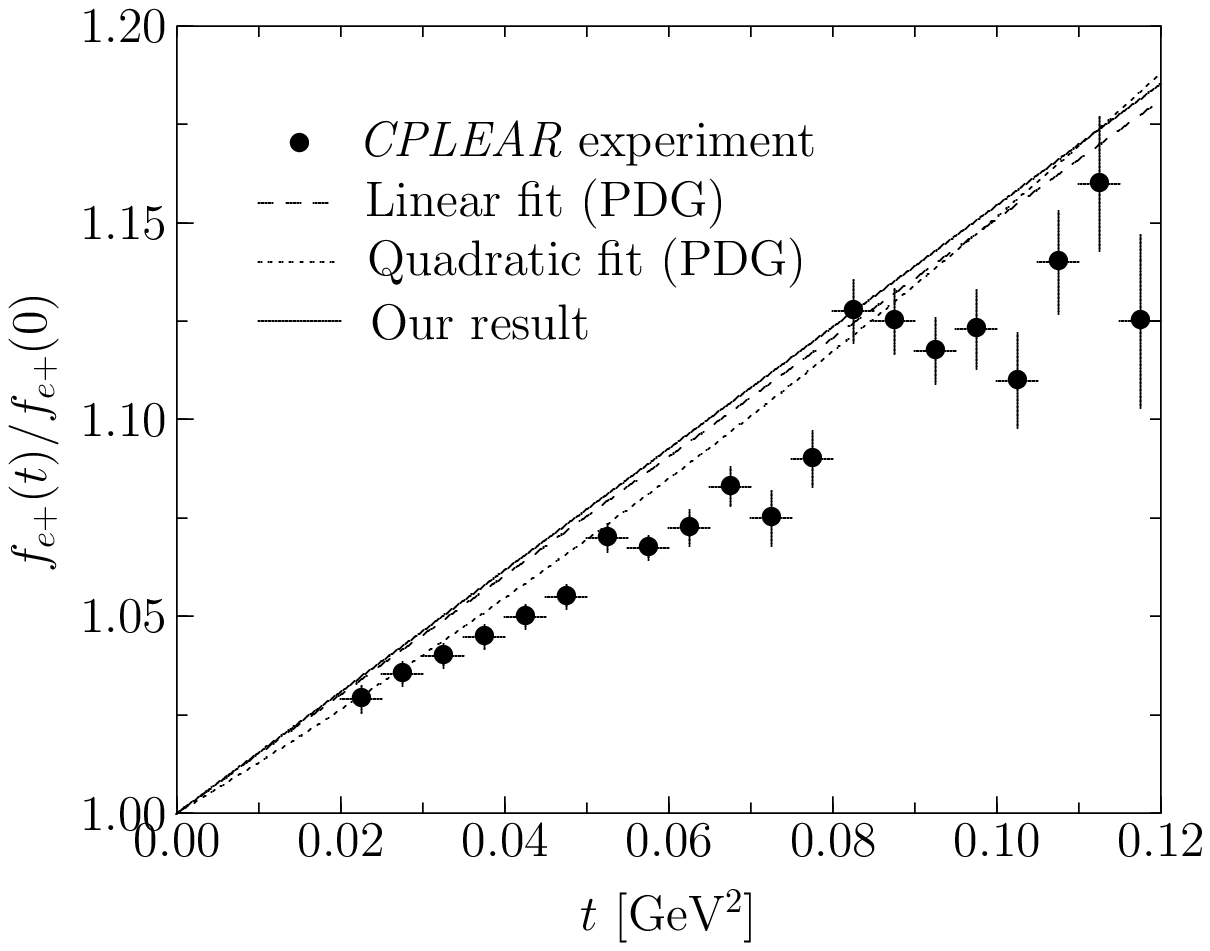}
\includegraphics[width=7.55cm]{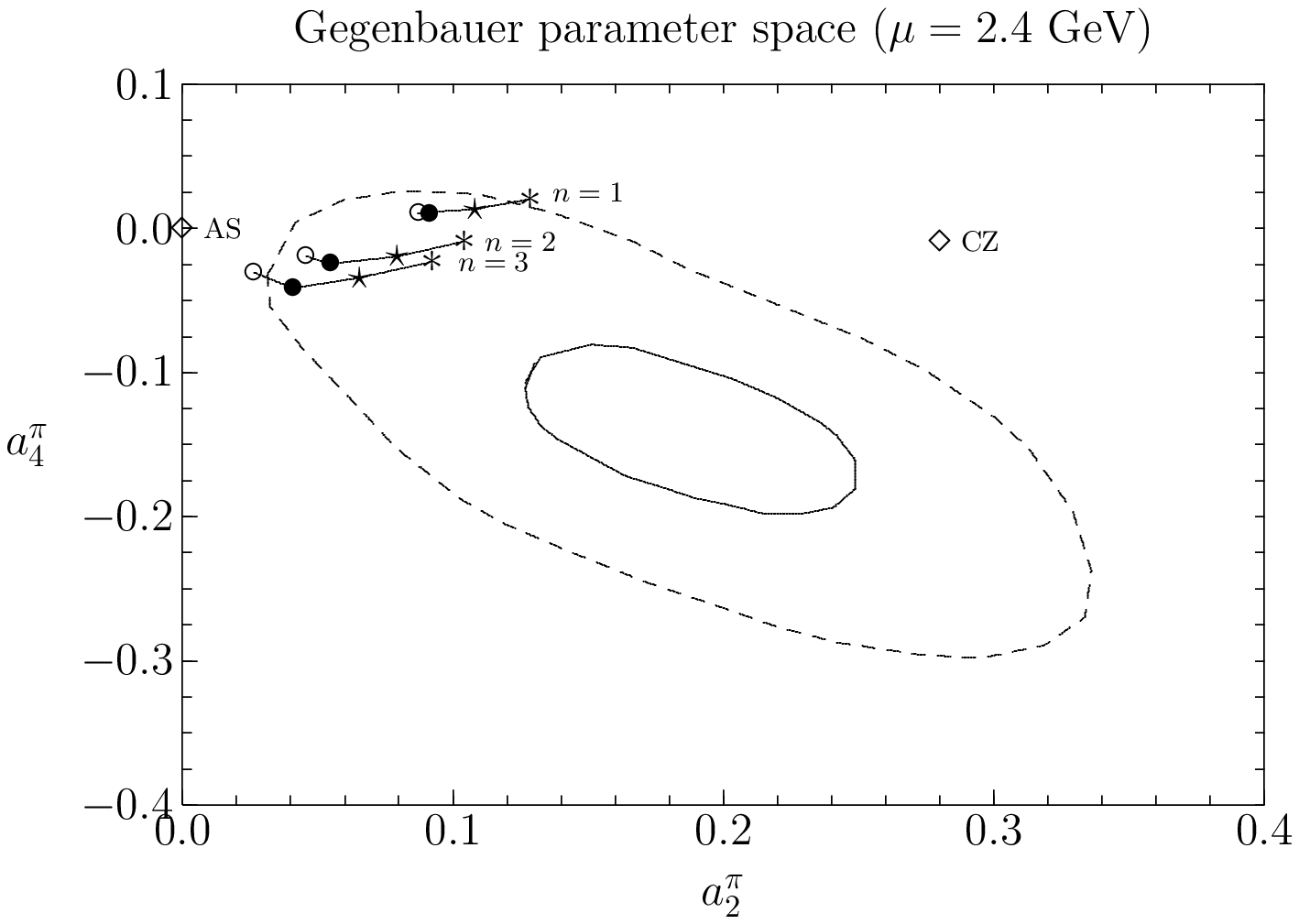}
\vspace*{8pt}
\caption{The results of the semileptonic form factors of
kaon~\protect\label{fig2}} 
\end{figure}
\section{Summary and conclusion}
In the present talk, we briefly reviewed recent investigations on the
structures of the pion and kaon from the instanton vacuum.  Without
adjusting any parameter, almost all results are in good agreement with
the experimental data.
\section*{Acknowledgments}
The authors are grateful to the organizers for the 4th Asia-Pacific
Conference on Few-Body Problems in Physics 2008 (APFB08). 
The present work is supported by the Korea Research Foundation Grant 
funded by the Korean Government(MOEHRD) (KRF-2006-312-C00507).
S.i.N. was partially supported by the Grant for Scientific Research
(Priority Area No.17070002 and No.20028005) from the Ministry of
Education, Culture, Science and Technology (MEXT) of Japan.

\end{document}